\documentclass[aps,prl,twocolumn,groupeaddress]{revtex4-1}

\usepackage{amsmath}
\usepackage{textcomp}
\usepackage{amsfonts}
\usepackage{amssymb}
\usepackage{inputenc}
\usepackage{fixltx2e}
\usepackage{graphicx}
\usepackage{color}
\usepackage{placeins}
\usepackage{natbib}
\usepackage{url}
\usepackage{textcase}
\usepackage{bm}

\begin{document}

\title{   Energy-dependent spatial texturing of the charge order in 1\textit{T}-Cu$_x$TiSe$_2$}

\author{M. Spera}

\author{A. Scarfato}

\author{E. Giannini}

\author{Ch. Renner}
 \email{Corresponding author. \\ christoph.renner@unige.ch \\}
 \affiliation{Department of Quantum Matter Physics, University of Geneva, 24 Quai Ernest-Ansermet, CH-1211 Geneva 4, Switzerland}

\date{\today}

\begin{abstract}

We report a detailed study of the microscopic effects of Cu intercalation on the charge density wave (CDW) in 1\textit{T}-Cu$_x$TiSe$_2$. Scanning tunneling microscopy and spectroscopy (STM/STS) reveal a unique, Cu driven spatial texturing of the charge ordered phase, with the appearance of energy dependent CDW patches and sharp $\pi$-phase shift domain walls ($\pi$DWs). The energy and doping dependencies of the patchwork are directly linked to the inhomogeneous potential landscape due to the Cu intercalants. They imply a CDW gap with unusual features, including a large amplitude, the opening below the Fermi level and a shift to higher binding energy with electron doping. Unlike the patchwork, the $\pi$DWs occur independently of the intercalated Cu distribution. They remain atomically sharp throughout the investigated phase diagram and occur both in superconducting and non-superconducting specimen. These results provide unique atomic-scale insight on the CDW ground state, questioning the existence of incommensurate CDW domain walls and contributing to understand its formation mechanism and interplay with superconductivity.

\end{abstract}


\maketitle

\section{Introduction}
Charge density waves (CDWs) are the focus of renewed interest, motivated in particular by a series of recent experiments addressing the interplay between charge order and superconductivity in a range of materials \cite{Ghiringhelli2012, Chang2012, Blanco-Canosa2013, Comin2014, McElroy2005, Wu2013, Tranquada1995, Hu2014, Tranquada1995, Fausti2011}. Furthermore, CDWs are still lacking a detailed understanding.  The gap amplitude remains matter of intense debate and robust evidence for nesting at the Fermi level, the preferred mechanism, is lacking in most CDW systems. Here, we address both issues in a detailed scanning tunneling microscopy and spectroscopy study of copper intercalated 1\textit{T}-TiSe$_\text{2}$.

Pristine TiSe$_\text{2}$ undergoes a phase transition into a commensurate $2a\times2b\times2c$ CDW at $T_{CDW} = 200 \text{ K}$, where $a$, $b$, and $c$ are the lattice parameters \cite{DiSalvo1976}. Among the main formation mechanisms proposed by theory are a Jahn-Teller distortion \cite{Hughes1977} and an excitonic coupling \cite{Wilson1969}. More recent studies propose a combination of these two mechanisms in a so-called indirect JT transition \cite{Kidd2002, Wezel2010} or a phonon-mediated excitonic interaction \cite{Phan2013}. Experiments suggest mixed contributions from both phonons and excitons, the former driving the CDW transition and the latter promoting long-range correlations \cite{Novello2017, Hildebrand2016, Porer2014}. Recently, signatures of an electronic soft mode at $\mathbf{q}_{CDW}$ have been detected via momentum resolved electron energy loss spectroscopy, consistent with the presence of an excitonic condensate in pristine TiSe$_\text{2}$ \cite{Kogar2017EELS}.

Cu$_x$TiSe$_2$ becomes superconducting above $x = 0.04$, with a maximum transition temperature $T_c = 4.1$ K near $x = 0.08$ \cite{Morosan2006}.
Transport measurements suggest that the CDW is progressively suppressed with increasing intercalated Cu content to ultimately vanish near the edge of the superconducting dome in the doping-temperature phase diagram. This interdependence has been interpreted as a competition between these two macroscopic quantum phases. However, recent transport and diffraction experiments hint at a more complex response of the CDW to Cu intercalation \cite{Kogar2017}, to ionic liquid gating \cite{Li2016} and to external pressure \cite{Joe2014}.
Among recent proposals is an incommensurate CDW phase developing above the superconducting dome for  $0.04 < x < 0.10$, that forms the breeding ground of the superconducting transition. In the present letter, we focus on the impact of Cu on the CDW phase for Cu concentration in the range $0\leq x< 0.07$. Our study gives a microscopic real space insight on the proposed CDW incommensurability, on its formation mechanism and interplay with superconductivity.

\section{Methods}
Single crystals of 1\textit{T}-Cu$_x$TiSe$_\text{2}$ with $0\leq x < 0.07$ were grown via iodine vapour transport. A stoichiometric mixture of titanium and selenium with selected amounts of copper was sealed in a quartz ampoule under high vacuum. The obtained crystals were annealed for one week at different temperatures depending on the final stoichiometry:  650 $^{\circ}$C for pristine ($x=0$) samples and 830 $^{\circ}$C for Cu intercalated samples ($x>0$). Scanning tunneling microscopy and spectroscopy (STM and STS) measurements were carried out with a SPECS Joule-Thomson STM with a base temperature of 1.2 K and a base pressure better than $1\times10^{-10}$ mbar. Indicated bias voltages refer to the sample bias. Tips were mechanically cut from a PtIr wire and conditioned in-situ on a Ag(111) single crystal. The samples were cleaved in-situ in ultra-high vacuum at room temperature shortly before mounting them on the STM head. STS measurements were performed using a Lock-in technique with a 3.54 mV rms bias modulation at $413.7 \text{ Hz}$, perfectly appropriate to detect spectral features in the $10$ mV range.

 \begin{figure}
 \includegraphics{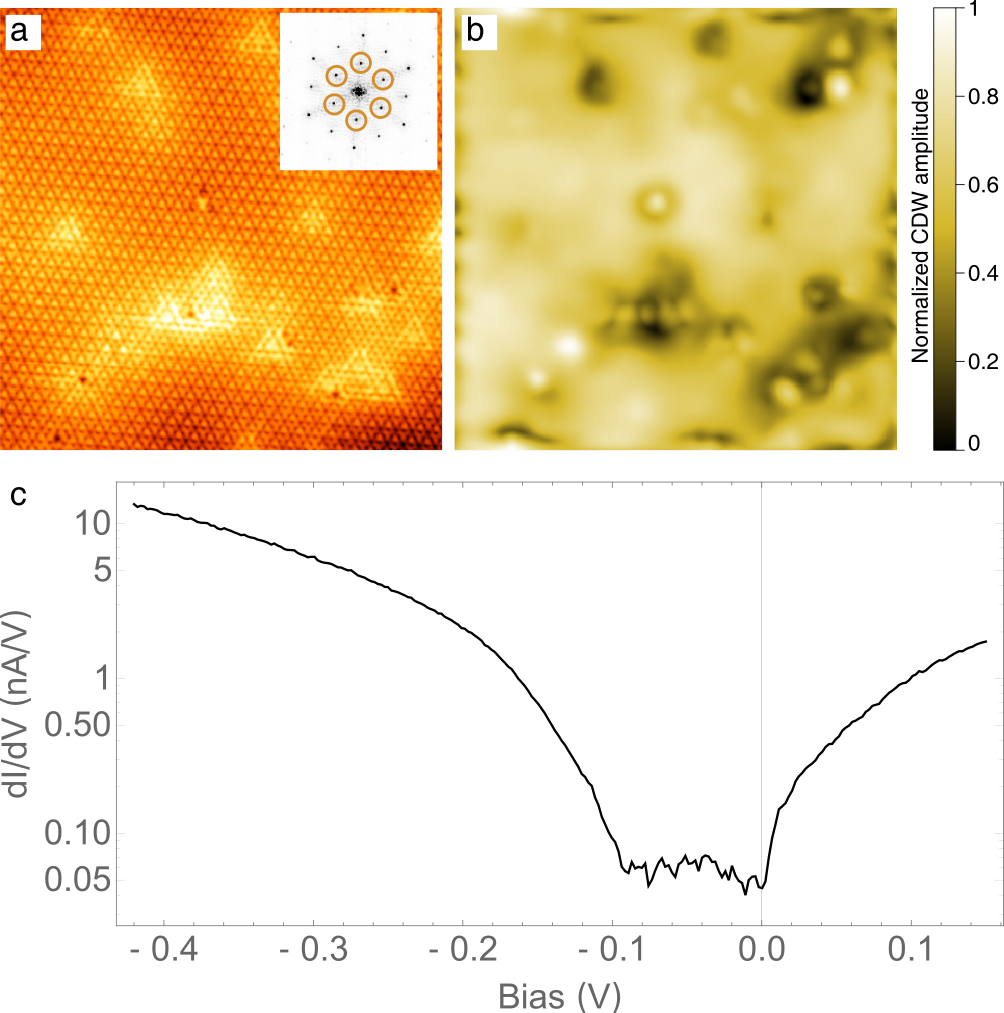}
 \caption{\label{Fig:0} (a) $19\times 19$ nm$^2$ STM micrograph of pristine TiSe$_2$  ($T=1.2$ K, $V=150$ mV, $I=100$ pA). Inset shows the corresponding FFT, where the $\mathbf{q}_{CDW}$ spots are highlighted by orange circles. (b) CDW amplitude map extracted from FFT filtering of panel (a), showing a uniform, long-range ordered CDW whose amplitude is vanishing near defects. (c) Differential tunneling conductance spectrum of TiSe$_2$ at 1.2~K on a  semi-logarithmic scale, showing a suppression of spectral weight below the Fermi level, consistent with ARPES measurements \cite{Rowher2014}.  Lock-in parameters: $V_{AC}^{rms} =3.54 \text{ mV}$, $f = 413.7 \text{ Hz}$.}
 \end{figure}

\begin{figure*}
 \includegraphics[width=\textwidth]{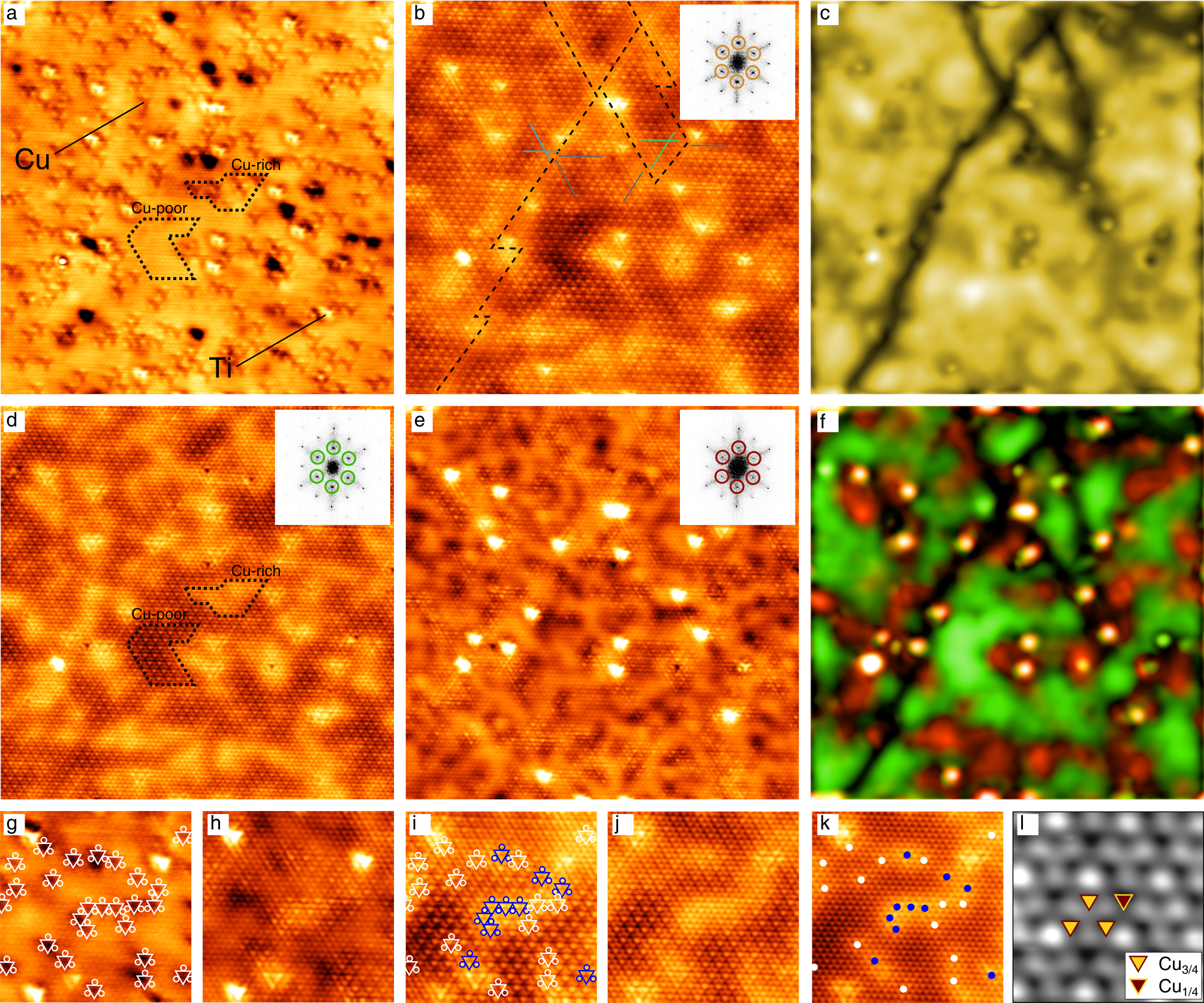}
 \caption{(color online) (a-f) $28 \times 28$ nm$^2$ STM images and CDW amplitude maps of the same Cu$_{0.02}$TiSe$_2$ surface region. (a) At V = -1.2 V, STM reveals intercalated Cu and Ti atoms (arrows). (b) At V = - 150 mV, CDW domains separated by narrow $\pi$-phase shift domain walls (dashed lines) are resolved over the entire surface. The solid lines are a guide to the eye following the CDW maxima to highlight the $\pi$ shift. (c) CDW amplitude map of panel (b), where the dark regions correspond to zero amplitude along the $\pi$DWs.
(d) The CDW appears suppressed in Cu-rich regions at $V = +150 \text{ mV}$ whereas in (e) it is suppressed in the perfectly complementary Cu-poor regions at $V = -350 \text{ mV}$. The insets show the FFT of each image, with the CDW spots highlighted by circles. (f) RGB sum of the CDW amplitude of panel (d) in green and panel (e) in red.
 (g-k) $11\times11$ nm$^2$ STM images of another region at (g) $V = -800 \text{ mV}$,  (h) $V = -300 \text{ mV}$, (i) $V = -150 \text{ mV}$,  (j) $V = +100 \text{ mV}$,  (k) $V = +300 \text{ mV}$ to show the Cu atom position relative to the CDW lattice (white: Cu$_{3/4}$ sites, blue: Cu$_{1/4}$ sites). Note the absence of CDW contrast at $V = +300 \text{ mV}$ in panel (k). A more detailed description is given in Supplementary Fig. 3 \cite{SM}.  (l) 1.8x1.8 nm$^2$ high resolution STM micrograph of Cu$_\text{0.02}$TiSe$_\text{2}$ at $150$ mV and $1.2$ K. Triangles depict the four possible octahedral Cu intercalation sites.   \label{Fig:1}}
 \end{figure*}

\section{Results}
Fig. \ref{Fig:0}(a) shows a STM micrograph of the in-situ cleaved surface of pristine 1\textit{T}-TiSe$_\text{2}$ at 1.2 K. A clear in-plane $2a\times 2b$ CDW modulation is resolved in real space and in $\mathbf{k}$ space, with a well-defined $\mathbf{q}_{CDW}$ superlattice outlined by orange circles in the Fast Fourier Transform (FFT) shown in the inset of Fig. \ref{Fig:0}(a).
Fig.~\ref{Fig:0}(b) is a map of the CDW modulation amplitude extracted from the FFT using the method described in Ref. \cite{Okamoto2015}. 
The amplitude is uniform over the whole surface investigated, except near defects (mostly Ti self doping) where it vanishes (dark color).
The intrinsic defect concentration is small enough not to perturb the long range order of the CDW \cite{Hildebrand2016}.
Tunneling spectroscopy (Fig. \ref{Fig:0}(c)) shows a loss of spectral weight in a finite energy window below the Fermi level. We associate this reduced local density of states (LDOS), whose energy range of $\sim 100$ meV below the Fermi level is consistent with photoemission spectra \cite{Rowher2014}, with the opening of the CDW gap. 

Intercalated Cu drastically affects the CDW, even at small concentrations of $x=0.02$ as shown in Fig. \ref{Fig:1}.
Confirming the implication of Cu in the observed CDW alterations is challenging, because Cu and CDW cannot be observed simultaneously. Indeed, Cu atoms can only be resolved at negative bias voltages below -800 mV  (Fig. \ref{Fig:1}(a)) \cite{Novello2017, Yan2017}, while CDW contrast is achieved at smaller bias voltages, within a few hundred millielectronvolts of the CDW gap (Figs.~\ref{Fig:1}(b), (d), (e), (g-k)) \cite{Novello2015, SM}. To align images taken at such different biases, we use well-documented fingerprints of atomic defects visible at all biases, in particular intercalated Ti \cite{Novello2015, Hildebrand2014}.

One of the striking features of the CDW images of copper intercalated TiSe$_\text{2}$ is an inhomogeneous electronic background (Fig. \ref{Fig:1}). We find this inhomogeneity is directly linked to intercalated Cu atoms which tend to cluster. This is best seen in high-resolution STM images (Fig.~\ref{Fig:1}(a), V=-1.2 V), revealing nanometre-scale regions where Cu is accumulating (identified as Cu-rich regions) and other regions with no copper atoms (Cu-poor regions). Note that on average, the number of intercalated Cu atoms resolved in large-scale images is in agreement with the nominal Cu doping (see Supplementary Information \cite{SM}). At positive imaging bias, Cu-rich (Cu-poor) regions appear brighter (darker) - one of each is pointed out for reference in Figs. \ref{Fig:1}(a) and \ref{Fig:1}(d).

\begin{figure}
 \includegraphics{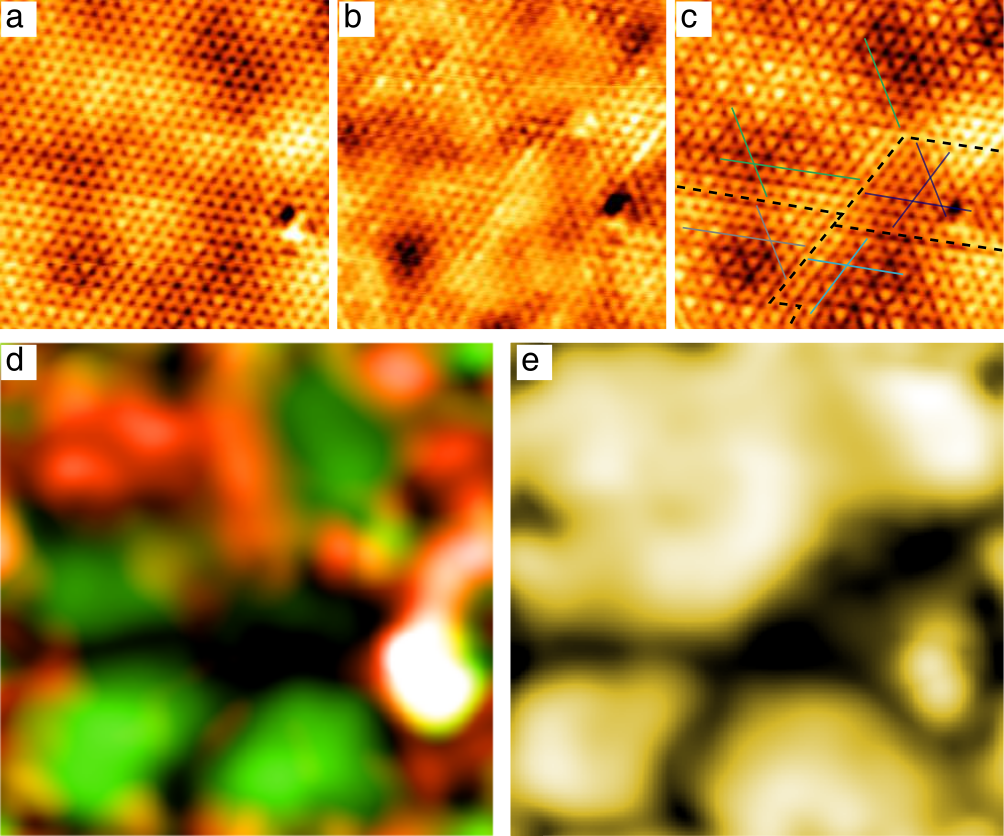}
 \caption{\label{Fig:3} (color online). $7.4\times 7.4$ nm$^2$ STM micrographs of Cu$_{0.06}$TiSe$_2$ at $T=5 $ K. The CDW is more disturbed than in the $x = 0.02$ crystal, but we still find it is suppressed in (a) Cu-rich regions at V = +150 mV and (b) in Cu-poor regions at V = -400 mV.  (c)  At a negative bias inside the gap  ($V = -150$ mV), the CDW is resolved over the entire image, except for a weaker modulation along the $\pi$-phase shift domain walls (dashed lines). The solid lines are guides to the eye following the CDW maxima to highlight the $\pi$ shift. (d) RGB sum of the CDW amplitude of panel (a) in green and panel (b) in red. (e) CDW amplitude map of panel (c), showing a finite CDW amplitude everywhere except along the $\pi$DWs, black regions where it drops to zero.  A more complete data set is given in Supplementary Fig. 4 \cite{SM}.}
 \end{figure}

Positive bias imaging shows a clear CDW modulation in the Cu-poor regions while it is strongly suppressed or even absent in the Cu-rich regions (Fig. \ref{Fig:1}(d)). Such interdependence would be expected for a CDW that is competing with superconductivity appearing above $x = 0.04$ Cu content \cite{Morosan2006}. 
However, a remarkably different and surprising picture emerges when imaging the same region at a negative bias of  $-350$ mV. STM now reveals a strong CDW amplitude in the Cu-rich regions and no CDW contrast in the Cu-poor ones (Fig.~\ref{Fig:1}(e)).  Remarkably, the CDW patches imaged at positive and negative sample bias in Figs. \ref{Fig:1}(d) and \ref{Fig:1}(e) span perfectly complementary areas on the sample surface.
This observation is averred in Fig. \ref{Fig:1}(f) where we plot the RGB sum of the local CDW amplitude of Fig. \ref{Fig:1}(d) expressed in green and that of Fig. \ref{Fig:1}(e) expressed in red. The absence of any yellow region (red+green) confirms the absence of overlapping regions in the positive and negative bias range discussed above. 

The analysis of Figs. \ref{Fig:1}(d)-(f) suggests that the entire surface is supporting a CDW, independent of the intercalated Cu distribution. Indeed, imaging the same region at a smaller negative bias ($-200$ mV) reveals a $2\times2$ CDW over the entire crystal surface (Fig. \ref{Fig:1}(b)), with a uniform amplitude, apart from sharp dark lines where it is suppressed, (Fig. \ref{Fig:1}(c)). 
The latter, identified by dashed lines in Fig. \ref{Fig:1}(b), correspond to $\pi$-phase shift domain walls ($\pi$DWs), which break the long-range order of the CDW with respect to the pristine
crystal.

The STM micrographs discussed above suggest that Cu intercalation affects the long range $2\times2$ commensurate CDW observed in the $ab$-plane of pristine crystals \cite{Hildebrand2014, Novello2015} in two ways: it induces a striking energy dependent patchwork of CDW regions (Figs. \ref{Fig:1}(d), (e), (f)) and promotes the formation of $\pi$-phase shift domain walls (Figs. \ref{Fig:1}(b), (c)).
Note the absence of contrast inversion expected for a standard electron-hole symmetric CDW \cite{Dai2014}. Indeed, the CDW maxima in the STM micrograph in Fig. \ref{Fig:1} remain pinned to the same atomic site, independent of bias voltage.

We observe an identical response of the CDW for all Cu concentrations considered, including superconducting crystals with $x>0.04$.  The 5 K data presented in Fig. \ref{Fig:3} are from a $x = 0.06$ superconducting sample with $T_c~=~3.1 \text{ K}$ as verified by resistivity and magnetic susceptibility. The prominent features are the same as for the non-superconducting $x = 0.02$ crystal presented in Fig.~\ref{Fig:1}, although with smaller, less defined CDW patches and more $\pi$DWs due to the increased Cu content \cite{SM}. Most remarkably, the $5$ K CDW pattern remains unchanged when cooling the sample to 1.2 K, deep into the superconducting phase (see Supplementary Fig. 5 \cite{SM}).

\section{Discussion}

Real space STM images (Fig. 2) reveal that intercalating Cu does not destroy the CDW, even for $x$ deep inside the superconducting dome (Fig. \ref{Fig:3} and \cite{SM}). Cu atoms intercalate on two inequivalent octahedral sites in the van der Waals gap with respect to the CDW modulation (Fig. \ref{Fig:1}(l)), with three Cu$_{3/4}$ sites available for every Cu$_{1/4}$ site. The experimental occupation of these sites in large scale images is of order 73\% and 27\%, respectively (see Supplementary Fig. 1 \cite{SM}). This is in good agreement with their expected abundances, indicating Cu is not detrimental to the CDW. Note this is different from Ti intercalation which goes on the same lattice site in the vdW gap \cite{Novello2015}: Ti$_{3/4}$ has been found to distort the local symmetry with the CDW adapting to minimize the associated energy cost by optimizing the occupation of the more favourable undistorted Ti$_{1/4}$ site \cite{Hildebrand2017}.

While the CDW amplitude remains finite for large $x$ in Cu$_x$TiSe$_2$, Cu does affect its long-range order via the formation of $\pi$DWs. The position of the $\pi$DWs is spatially uncorrelated with intercalated Cu sites, but their number and density are increasing with Cu content. We associate the loss of long range charge order with the exciton melting due to the increased metallicity of the samples with Cu doping. As the Fermi level moves to higher energy with Cu band doping \cite{Zhao2007}, the exciton condensate melts as the shift becomes of the order of the exciton binding energy (about 17 meV according to ref. \cite{Pillo2000}). This effect can account for the decreasing transition temperature observed in transport measurements \cite{Morosan2006}. However, the persistence of short range CDW domains shows that the ordered ground state develops independent of excitons. This idea has previously been proposed and discussed based on momentum transfer \cite{Porer2014} and real space imaging \cite{Novello2017, Hildebrand2017} experiments. These findings invalidate both a Fermi surface nesting scenario and a purely electronic CDW formation mechanism \cite{Wezel2010, Novello2017, Porer2014}.

$\pi$DWs have been associated with a slight incommensuration in the CDW \cite{Yan2017}, and incommensurate CDW (ICDW) domain walls have been proposed to promote superconductivity \cite{Li2016, Joe2014, Kogar2017}. However, according to our STM experiments, $\pi$DWs  appear at much lower Cu concentration than superconductivity (Fig. \ref{Fig:1}), and they have been found to proliferate in  Ti intercalated TiSe$_2$ \cite{Hildebrand2016} which does not become superconducting. The domain walls we resolve by STM are atomically sharp boundaries where the CDW amplitude drops to zero, independent of Cu content (black contrast in Figs. \ref{Fig:1}(c) and \ref{Fig:3}(e)). They separate perfectly commensurate $2\times 2$ CDW domains (yellow amplitudes in Figs. \ref{Fig:1}(c) and \ref{Fig:3}(e)) without any sign of an incommensurate charge order. These experimental observations invalidate the proposal that incommensurate domain walls are driving the emergence of superconductivity in Cu$_x$TiSe$_2$ \cite{Li2016, Joe2014, Kogar2017}.

 \begin{figure}
 \includegraphics{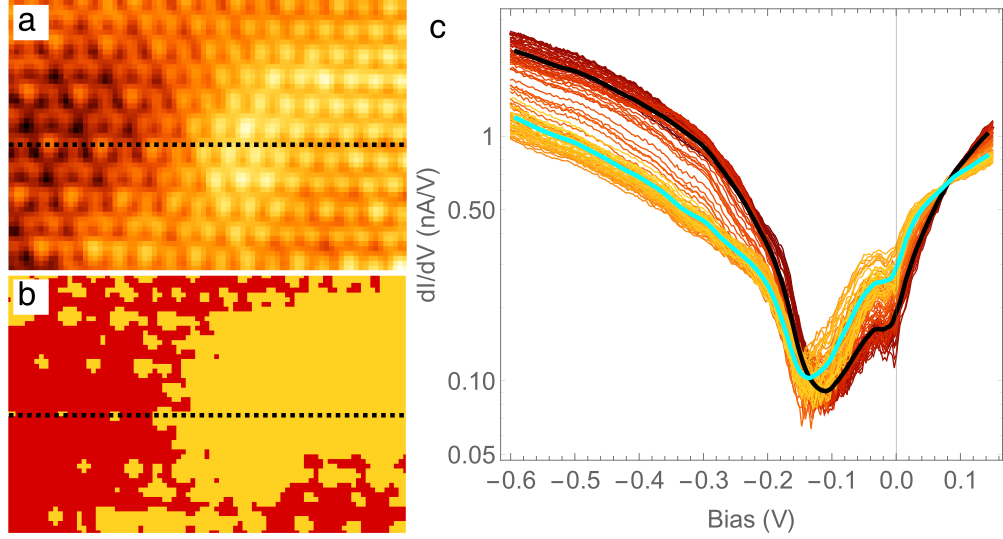}
 \caption{\label{Fig:2} (color online). (a) Topography acquired during a differential conductance map with $85 \times 55$ spectra over a $5.5\times3.3$ nm$^2$ surface of Cu$_{0.02}$TiSe$_\text{2}$ at $T = 1.2$ K. (b) k-mean clustering map of all the acquired spectra, showing a clear separation between Cu-poor and Cu-rich regions. (c) Differential conductance spectra in semi-logarithmic scale along the line in (a) and (b), showing a position-dependent band shift from the Cu poor region on the left-hand side of panel (a) to the Cu-rich region in on the right-hand side of panel (a). The light blue and black curves correspond to the k-means average of these curves in two families.
 Setpoint: $V_{set} = 150 \text{ mV}$, $I_{set} = 100 \text{ pA}$.  Lock-in parameters: $V_{AC}^{rms} =3.54 \text{ mV}$, $f = 413.7 \text{ Hz}$. }
 \end{figure}

We now focus on the energy dependent CDW patchwork developing upon Cu intercalation. Contrary to the $\pi$DWs which form independently of the location of the Cu atoms, the CDW patchwork is closely linked to the spatial Cu distribution. To understand this patchwork, we examine the spatial dependence of the tunneling spectra on a crystal with $x=0.02$. On average, compared to pristine TiSe$_2$, the spectral features shift towards higher binding energy, consistent with the electron band dopant character of Cu reported earlier \cite{Zhao2007, Novello2017}. A closer inspection of  STS data over a $5.5 \times 3.3$~nm$^2$ area straddling a Cu-poor and a Cu-rich region (Fig.~\ref{Fig:2}(a)) shows that this shift is sensitive to the local Cu content, with a slightly larger shift in Cu-rich compared to Cu-poor regions (Fig.~\ref{Fig:2}(c)). Bunching all the spectra into two families using a k-means clustering analysis shows that more (less) shifted spectra are associated with Cu-rich (Cu-poor) regions depicted in yellow (red) on Fig.~\ref{Fig:2}(b).

\begin{figure}[t]
 \includegraphics{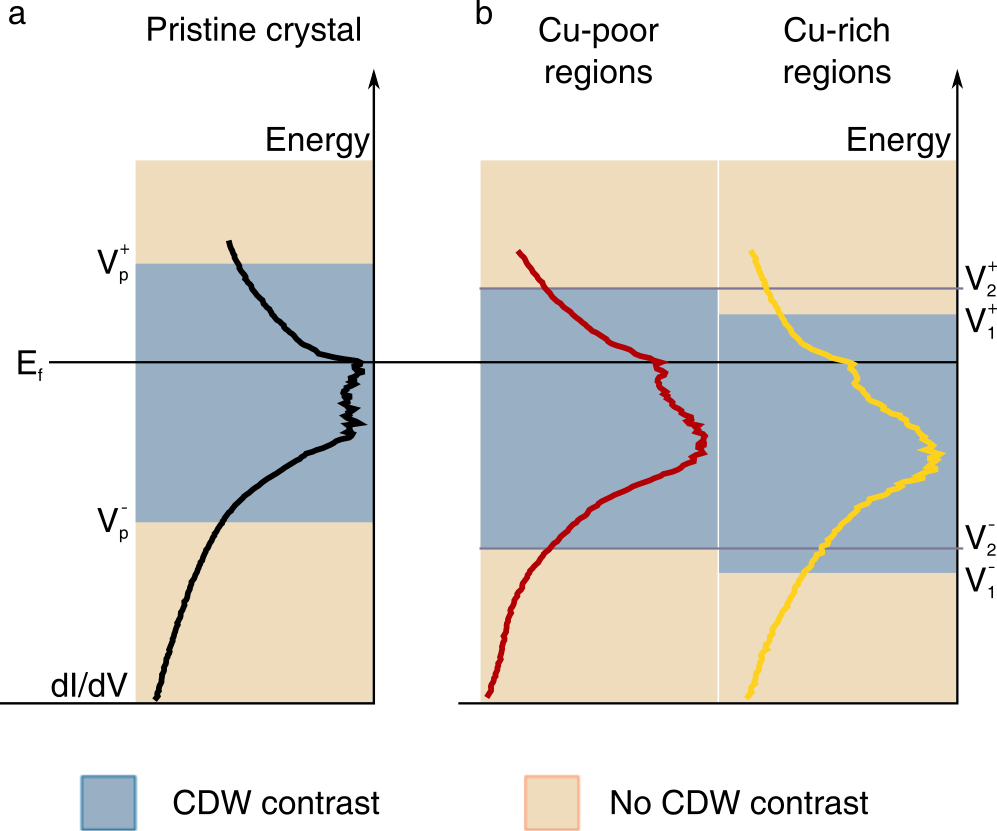}
 \caption{\label{Fig:4} (color online). Schematic depiction of the CDW imaging conditions in Cu$_x$TiSe$_2$. Solid lines show typical local tunneling spectra of (a) a pristine ($x = 0$) and (b) a Cu intercalated ($x = 0.02$) crystal.  The blue rectangle represents the limited bias range  $V^-$ to $V^+$ where STM imaging yields CDW contrast. This CDW imaging window shifts to higher binding energies depending on the local Cu content. A direct consequence is different energy ranges where the CDW can be resolved in Cu-rich ($V_1^-$ to $V_1^+$) and Cu-poor ($V_2^-$ to $V_2^+$) regions.}
 \end{figure}

Now that we have established a spatial correlation between the shift of spectral features and the local Cu content, we show that the observed Cu driven band shift is directly responsible for the energy dependent CDW patchwork displayed in Figs.~\ref{Fig:1} and~\ref{Fig:3}. Indeed, the tunneling bias voltage has to be set in a specific energy range around the gap feature to achieve CDW contrast. This is schematically shown in Fig.~\ref{Fig:4}(a) for a pristine crystal, where CDW contrast is achieved between $V_p^-$ and $V_p^+$, and further visualised in Supplementary Fig. 2. The imaging bias range is asymmetric with respect to $E_F$ and in apparent relation to the gap that opens below the Fermi level.

With increasing local Cu content, the spectral features near the Fermi energy shift to higher binding energies. Assuming the CDW imaging bias range shifts alongside the spectral features, and in particular alongside the reduced LDOS below $E_F$, offers  a straightforward explanation for the energy dependent and bias complementary CDW patchwork. Explicitly, setting the tunneling bias between $V_2^-$ and $V_2^+$ allows to resolve the CDW best in Cu-poor regions (Figs. \ref{Fig:1}(d) and \ref{Fig:3}(a)), whereas between $V_1^-$ and $V_1^+$ it is best seen in Cu-rich regions (Figs. \ref{Fig:1}(e) and \ref{Fig:3}(b)). Since these two energy ranges do not completely overlap, there are two finite energy windows where the CDW can only be resolved either in Cu-poor ($V_1^+$ to $V_2^+$) or in Cu-rich regions ($V_1^-$ to $V_2^-$). Setting the bias voltage between $V_2^-$ and $V_1^+$, STM resolves $2 \times 2$ CDW domains separated by $\pi$DWs over the entire surface with no apparent patchwork (Figs. \ref{Fig:1}(b) and \ref{Fig:3}(c)), while no CDW contrast is obtained below $V_1^-$ and above $V_2^+$ (Figs. \ref{Fig:1}(a) and \ref{Fig:1}(m)). These results are further visualised with the aid of the FFT filtering method in Supplementary Fig. 6 \cite{SM}.
 Note that Ti, which is also an electron donor intercalating on the same lattice site  as Cu \cite{Novello2015}, has been associated with $\pi$DWs \cite{Hildebrand2016}  but does not trigger an energy dependent CDW patchwork like the one we find with Cu (see Supplementary Fig. 2 \cite{SM}).

The energy dependent CDW patchwork uncovered here suggests that the electronic states involved in the CDW phase are not centered on the Fermi level and shift to higher binding energies when the local carrier density is increased. Its non-symmetric energy dependence with respect to the Fermi level and its evolution with increasing Cu content provides additional evidence to the tunneling spectroscopy that the CDW gap is opening below $E_F$ and is shifting to higher binding energies when the system turns more metallic upon Cu intercalation. While the patchwork provides evidence for a link between the CDW phase and the measured gap, its unusually large amplitude of about $100$ meV, consistent with ARPES, calls for further investigations.

Finally, the CDW patchwork sheds a different light on a recent study of ion-liquid gating doped TiSe$_2$ films \cite{Li2016}. Ion liquid gating yields a spatially inhomogeneous carrier doping \cite{Costanzo2016, Petach2017} and we expect the associated non-uniform potential landscape to promote energy dependent CDW patches similar to those reported here. The concomitant potential landscape and spatial variations of the carrier density at the Fermi level can account for the reported Little-Parks effect, as the supercurrents would move in loops within domains defined by these charge inhomogeneities, independent of the local charge order. While the domain size is compatible with the superconducting coherence length in these samples  \cite{Kacmarcik2013, Hillier2010,  Qian2007, Zaberchik2010, Husanikova2011}, the CDW patches uncovered here are too small to confine superconductivity. We do thus not expect to observe the above Little-Parks effect in superconducting Cu$_x$TiSe$_2$ crystals.

\section{Conclusions}

We present a detailed scanning tunneling microscopy investigation of the CDW response to Cu intercalation in 1\textit{T}-Cu$_x$TiSe$_2$. We identify two mechanisms that can account for the fading of the CDW signal reported in bulk experiments. First, the loss of long range order via the proliferation of $\pi$DWs with increasing copper content. 
Second, a striking energy dependent patchwork driven by the inhomogeneous energy landscape due to the random distribution of intercalated Cu atoms. 
 The vanishing CDW amplitude at the $\pi$DWs and their atomic-scale width are not compatible with incommensurate CDW domains proposed in recent publications \cite{Kogar2017, Li2016}. 
The patchwork provides direct evidence that the CDW gap in 1\textit{T}-Cu$_x$TiSe$_2$ is opening below the Fermi level and is shifting to higher binding energies with increasing Cu content.  Interestingly, STM finds the CDW to survive for Cu doping deep inside the superconducting dome, hinting at a possible coexistence of these two ground states. Our results show that the emergence of superconductivity is most likely not related to the weakening and decoherence of the CDW, but is promoted by the Cu driven band shift and subsequent enhanced density of states at the Fermi level.
We also find no evidence for incommensurate CDW domain walls which have been proposed to promote superconductivity.
Further insight into the interplay of charge order and superconductivity requires a detailed mapping of the superconducting gap, which is beyond the scope of this study.

\section{Acknowledgements}

This project was supported by the Swiss National Science Foundation through Div. II (grant 162517). We acknowledge stimulating discussions with D. R. Bowler, A. Morpurgo, Ch. Berthod, A. Pasztor, J. van Wezel, P. Aebi, B. Hildebrand and J. Lorenzana. We thank C. Barreteau for her help with the transport measurements, and G. Manfrini and A. Guipet for their technical assistance.

\bibliography{biblio_decoherence}{}
\bibliographystyle{apsrev4-1}

\end{document}